\documentclass[11pt,twoside]{article}

\usepackage{asp2014}

\aspSuppressVolSlug
\resetcounters

\begin{document}

\title{An Overview of the LSST Image Processing Pipelines}

\author{
    James~Bosch$^{1,2}$,
    Yusra~AlSayyad$^2$,
    Robert~Armstrong$^3$,
    Eric~Bellm$^4$,
    Hsin-Fang~Chiang$^5$,
    Siegfried~Eggl$^4$,
    Krzysztof~Findeisen$^4$,
    Merlin~Fisher-Levine$^6$,
    Leanne~P.~Guy$^6$,
    Augustin~Guyonnet$^7$,
    {\v{Z}}eljko~Ivezi{\'c}$^4$,
    Tim~Jenness$^6$,
    G{\'a}bor~Kov{\'a}cs$^4$,
    K.~Simon~Krughoff$^6$,
    Robert~H.~Lupton$^2$,
    Nate~B.~Lust$^2$,
    Lauren~A.~MacArthur$^2$,
    Joshua~Meyers$^3$,
    Fred~Moolekamp$^{2,8}$,
    Christopher~B.~Morrison$^4$,
    Timothy~D.~Morton$^{2,9}$,
    William~O'Mullane$^6$,
    John~K.~Parejko$^4$,
    Andr{\'e}s~A.~Plazas$^2$,
    Paul~A.~Price$^2$,
    Meredith~L.~Rawls$^4$,
    Sophie~L.~Reed$^2$,
    Pim~Schellart$^2$,
    Colin~T.~Slater$^4$,
    Ian~Sullivan$^4$,
    John.~D.~Swinbank$^4$,
    Dan~Taranu$^2$,
    Christopher~Z.~Waters$^2$,
    and
    W.~M.~Wood-Vasey$^{10}$
}
\affil{$^1$\email{jbosch@astro.princeton.edu}}
\affil{$^2$Princeton University, Princeton, NJ, U.S.A.}
\affil{$^3$Lawrence Livermore National Laboratory, Livermore, CA, U.S.A.}
\affil{$^4$University of Washington, Seattle, WA, U.S.A.}
\affil{$^5$National Center for Supercomputing Applications, Urbana, IL, U.S.A.}
\affil{$^6$LSST Project Management Office, Tucson, AZ, U.S.A.}
\affil{$^7$Harvard University, Cambridge, MA, U.S.A}
\affil{$^8$Rider University, Lawrenceville, NJ, U.S.A.}
\affil{$^9$University of Florida, Gainesville, FL, U.S.A.}
\affil{$^{10}$University of Pittsburgh, Pittsburgh, PA, U.S.A.}

\paperauthor{James~Bosch}{jbosch@astro.princeton.edu}{0000-0003-2759-5764}{Princeton University}{Department of Astrophysical Sciences}{Princeton}{NJ}{08544}{U.S.A.}
\paperauthor{Yusra~AlSayyad}{}{}{Princeton University}{Department of Astrophysical Sciences}{Princeton}{NJ}{08544}{U.S.A.}
\paperauthor{Robert~Armstrong}{}{}{Lawrence Livermore National Laboratory}{}{Livermore}{CA}{94551}{U.S.A}
\paperauthor{Eric~Bellm}{ecbellm@uw.edu}{0000-0001-8018-5348}{University of Washington}{Department of Astronomy}{Seattle}{WA}{98195}{U.S.A.}
\paperauthor{Hsin-Fang~Chiang}{hchiang2@illinois.edu}{0000-0002-1181-1621}{NCSA}{University of Illinois at Urbana-Champaign}{Urbana}{IL}{61801}{U.S.A.}
\paperauthor{Siegfried~Eggl}{}{}{University of Washington}{Department of Astronomy}{Seattle}{WA}{98195}{U.S.A.}
\paperauthor{Krzysztof~Findeisen}{kfindeis@uw.edu}{0000-0003-1898-5760}{University of Washington}{Department of Astronomy}{Seattle}{WA}{98195}{U.S.A.}
\paperauthor{Merlin~Fisher-Levine}{}{0000-0001-9440-8960}{LSST}{Data Management}{Tucson}{AZ}{85719}{U.S.A.}
\paperauthor{Leanne~P.~Guy}{lguy@lsst.org}{0000-0003-0800-875}{LSST}{Data Management}{Tucson}{AZ}{85719}{U.S.A.}
\paperauthor{Augustin~Guyonnet}{}{0000-0003-0879-1292}{Harvard University}{Department of Astronomy}{Cambridge}{MA}{02138}{U.S.A.}
\paperauthor{{\v{Z}}eljko~Ivezi{\'c}}{ivezic@astro.washington.edu}{}{University of Washington}{Department of Astronomy}{Seattle}{WA}{98195}{U.S.A.}
\paperauthor{Tim~Jenness}{tjenness@lsst.org}{0000-0001-5982-167X}{LSST}{Data Management}{Tucson}{AZ}{85719}{U.S.A.}
\paperauthor{G{\'a}bor~Kov{\'a}cs}{}{}{University of Washington}{Department of Astronomy}{Seattle}{WA}{98195}{U.S.A.}
\paperauthor{K.~Simon~Krughoff}{}{0000-0002-4410-7868}{LSST}{Data Management}{Tucson}{AZ}{85719}{U.S.A.}
\paperauthor{Robert~H.~Lupton}{rhl@astro.princeton.edu}{0000-0003-1666-0962}{Princeton University}{Department of Astrophysical Sciences}{Princeton}{NJ}{08544}{U.S.A.}
\paperauthor{Nate~B.~Lust}{}{0000-0002-4122-9384}{Princeton University}{Department of Astrophysical Sciences}{Princeton}{NJ}{08544}{U.S.A.}
\paperauthor{Lauren~A.~MacArthur}{lauren@astro.princeton.edu}{}{Princeton University}{Department of Astrophysical Sciences}{Princeton}{NJ}{08544}{U.S.A.}
\paperauthor{Joshua~Meyers}{}{}{Lawrence Livermore National Laboratory}{}{Livermore}{CA}{94551}{U.S.A}
\paperauthor{Fred~Moolekamp}{}{0000-0003-0093-4279}{Rider University}{Department of Chemistry, Biochemistry, and Physics}{Lawrenceville}{NJ}{08648}{U.S.A.}
\paperauthor{Christopher~B.~Morrison}{}{}{University of Washington}{Department of Astronomy}{Seattle}{WA}{98195}{U.S.A.}
\paperauthor{Timothy~D.~Morton}{}{0000-0002-8537-5711}{University of Florida}{Department of Astronomy}{Gainesville}{FL}{32611}{U.S.A.}
\paperauthor{William~O'Mullane}{womullan@lsst.org}{}{LSST}{Data Management}{Tucson}{AZ}{85719}{U.S.A.}
\paperauthor{John~K.~Parejko}{parejkoj@u.washington.edu}{}{University of Washington}{Department of Astronomy}{Seattle}{WA}{98195}{U.S.A.}
\paperauthor{Andr{\'e}s~A.~Plazas}{}{0000-0002-2598-0514}{Princeton University}{Department of Astrophysical Sciences}{Princeton}{NJ}{08544}{U.S.A.}
\paperauthor{Paul~A.~Price}{}{}{Princeton University}{Department of Astrophysical Sciences}{Princeton}{NJ}{08544}{U.S.A.}
\paperauthor{Meredith~L.~Rawls}{}{0000-0003-1305-7308}{University of Washington}{Department of Astronomy}{Seattle}{WA}{98195}{U.S.A.}
\paperauthor{Sophie~L.~Reed}{}{0000-0002-4422-0553}{Princeton University}{Department of Astrophysical Sciences}{Princeton}{NJ}{08544}{U.S.A.}
\paperauthor{Pim~Schellart}{}{}{Princeton University}{Department of Astrophysical Sciences}{Princeton}{NJ}{08544}{U.S.A.}
\paperauthor{Colin~T.~Slater}{ctslater@uw.edu}{0000-0002-0558-0521}{University of Washington}{Department of Astronomy}{Seattle}{WA}{98195}{U.S.A.}
\paperauthor{Ian~Sullivan}{sullii@uw.edu}{0000-0001-8708-251X}{University of Washington}{Department of Astronomy}{Seattle}{WA}{98195}{U.S.A.}
\paperauthor{John~D.~Swinbank}{}{0000-0001-9445-1846}{University of Washington}{Department of Astronomy}{Seattle}{WA}{98195}{U.S.A.}
\paperauthor{Dan~Taranu}{}{0000-0001-6268-1882}{Princeton University}{Department of Astrophysical Sciences}{Princeton}{NJ}{08544}{U.S.A.}
\paperauthor{Christopher~Z.~Waters}{}{0000-0003-1989-4879}{Princeton University}{Department of Astrophysical Sciences}{Princeton}{NJ}{08544}{U.S.A.}
\paperauthor{W.~M.~Wood-Vasey}{}{}{University of Pittsburgh}{Pittsburgh Particle Physics, Astrophysics, and Cosmology Center (PITT PACC), Physics and Astronomy Department}{Pittsburgh}{PA}{15260}{U.S.A.}


\begin{abstract}
The Large Synoptic Survey Telescope (LSST) is an ambitious astronomical survey with a similarly ambitious Data Management component.
Data Management for LSST includes processing on both nightly and yearly cadences to generate transient alerts, deep catalogs of the static sky, and forced photometry light-curves for billions of objects at hundreds of epochs, spanning at least a decade.
The algorithms running in these pipelines are individually sophisticated and interact in subtle ways.
This paper provides an overview of those pipelines, focusing more on those interactions than the details of any individual algorithm.
\end{abstract}

\section{Introduction}

Over the course of the 2020s, the Large Synoptic Survey Telescope \citep{2008arXiv0805.2366I} will produce a petabyte-scale astronomical dataset with an unprecedented combination of depth, area, and time-domain sensitivity across six optical\slash near-infrared bands.
The LSST project is much more than a telescope; it is first and foremost a survey, and one accompanied by an extensive data management effort.
LSST Data Management \citep{2017ASPC..512..279J,2018AAS...23136210O} is responsible for producing many different data products as well as providing services to archive and serve the data to the community.
This paper provides an overview of the pipelines and algorithms responsible for generating those data products, including high-level, ``science-ready'' catalogs.

The LSST processing pipelines encompass two major components.
The Prompt Processing\footnote{
    ``Prompt Processing'' is now preferred over the ``Alert Production'' terminology LSST has used in the past, reflecting the fact that these pipelines generate more than just alerts.
} pipelines will run in near-real-time, generating alerts for transient detections from image differencing within 60 seconds of their observation, as well as subsequent forced photometry and orbit updates for solar system objects over the course of the next 24 hours.
The Data Release pipelines will run on a yearly cadence\footnote{The first two data releases are currently planned to be only six months apart.}, producing a complete reprocessing of the full survey dataset each time.
The Data Release pipelines generate calibrated images, coadds, image differences, and catalogs of detections and measurements derived from all of these.

The catalogs also split into two general categories.
In the LSST nomenclature, \textit{object} catalogs have a single entry for each astrophysical object, which in general aggregates measurements from multiple observations.
\textit{Source} catalogs have different entries for each observation of an object.
Multiple catalogs in both categories are produced in both Prompt Processing and Data Release production.

The inclusion of image differencing in both the Data Release pipelines and the Prompt Processing pipelines is worth drawing attention to, as it highlights the fact that the algorithms developed by LSST cannot be cleanly split into Prompt and Data Release components.
Both groups of pipelines are built on top of a common algorithmic codebase and middleware system \citep[e.g.][]{P13-7_adassxxviii}, and each Data Release processing campaign will re-do almost everything done by the previous year's\footnote{
    This is a slight simplification of the timing; the set of raw observations included in an LSST Data Release is frozen before processing starts.
    That processing will take the better part of a year, so it may be more than a year before a particular observation first appears in a data release.
}
Prompt processing.

The LSST pipelines and algorithms have been in development for more than a decade; they are very much functional, but they are by no means complete.
This paper is intended to summarize their expected state in early operations.
For a much more thorough description of many of these algorithms in their current state, we refer the reader to \citet{2018PASJ...70S...5B}, which describes the processing of the Hyper Suprime-Cam Strategic Survey Program \citep{2018PASJ...70S...4A} using software derived from a recent version of the LSST codebase.
Project documents also provide additional information on planned LSST data products \citep[LSE-163;][]{LSE-163} and algorithms \citep[LDM-151;][]{LDM-151}.

\section{Goals and Philosophy}

Official survey pipelines like those being developed for LSST must be
designed with an extremely broad range of scientific use cases in mind.
This can actually make the goals and priorities of survey pipeline development quite different those of algorithm development in the pursuit of more specific science goals, despite similarities in methodology.

The outputs of an official survey pipeline are expected to act as proxies for the raw data; they should be as free of assumptions, filtering, and biases as possible.
Avoiding any kind of modeling is of course impossible: catalogs are themselves models of the sky as a set of discrete objects, built on top of models of the observatory and atmosphere.
Ideally survey pipelines employ multiple models, deferring the selection of the most appropriate model to downstream analyses.
For example, we measure the fluxes of each object under both the assumption that it is a point source and the assumption that it is a galaxy, rather than classifying it first and using that classification to determine how to photometer it.
Given the diversity of astrophysical objects and science goals, this naturally leads to a proliferation of measurements, which may seem at first to be a waste of processing time and storage, given that many astronomical objects can be securely classified to a degree that makes some measurements obviously inappropriate.
In practice, however, the computational and storage demands of a survey are driven by its faintest and most poorly-resolved populations, where classifications are rarely secure.

Multiple models are unfortunately not an option in the early stages of pipeline processing; we simply cannot afford to fit multiple \emph{image characterization} models, such as the point-spread function (PSF) or the sky background, given that utilizing different versions of those models in downstream processing would lead to a combinatorial explosion in the catalog size.
Instead, the quality of these models must meet the requirements of the most demanding downstream science, and in some cases, those requirements are in tension.
Astronomers interested in faint point sources, for example, prefer local background estimates even when this conflates sky backgrounds with smooth, low-surface-brightness features from nearby extended objects, while those interested in those low-surface-brightness objects require backgrounds to be measured only on very large spatial scales.
A survey pipeline that meets the requirements of both science cases must thus use a better overall background model than either science case would require independently.

Survey pipelines also typically operate under stringent computational and storage constraints, however.
To make the above ``deferred-model-selection'' and ``best-possible-image-characterization-models'' philosophies computationally tractable, survey pipelines such as those used in the SDSS \citep[\textit{Photo;}][]{2001ASPC..238..269L} and Pan-STARRS \citep[\textit{IPP;}][]{2016arXiv161205245W} traditionally have made two major simplifying assumptions:
\begin{itemize}
\item astronomical objects are sufficiently separated on the sky to be detected and well-measured independently (even if this involves an explicit \emph{deblending} step to further separate them);
\item calibrations such as the PSF and background can be characterized well enough that their uncertainties can be neglected when computing the overall errors on per-object measurements (even if their biases cannot be).
\end{itemize}

The applicability of these assumptions and trade-offs has largely been borne out in the science results.
Bayesian methods that relax these assumptions by jointly sampling or optimizing multi-object likelihoods have yet to produce results that improve on traditional pipelines in more than very limited respects, despite being orders of magnitude more computationally expensive \citep{2016arXiv161103404R,2013AJ....146....7B,2015ApJ...807...87S}; in many cases, they are still too expensive for fair comparisons to even be made.

This may change in the era of LSST, however, because the survey's depth comes with a dramatic increase in object density, and its size (and accompanying reduction in statistical errors) tightens requirements on systematic errors.
The SDSS deblending algorithm in particular has already been demonstrated to be inadequate at LSST depths \citep{2018PASJ...70S...5B}, and it remains unclear whether an improved algorithm with essentially the same philosophical approach can solve this problem.
Methods based more directly on joint or iterative fitting of multi-object likelihoods \citep[e.g.][]{2018A&C....24..129M,2012MNRAS.422..449B,2018ApJS..235...33D} probably have a role to play as well, but defining constraints or priors that are informative enough to ensure efficient convergence without biasing derived measurements is a significant challenge.

\section{Prompt Pipelines and Data Products}

\subsection{Single-Epoch Processing}

Processing of LSST science observations begins with Instrument Signature Removal (ISR), which includes basic detrending (flat-fielding, bias subtraction, fringe correction, etc), nonlinearity and crosstalk correction, and masking of bad and saturated pixels.
A full description of LSST's photometric and astrometric calibration plans is well beyond the scope of this paper, but it is worth noting that the flat applied here is a fairly sophisticated construct, derived from data from many different sources, including:
\begin{itemize}
\item traditional dome flats, to constrain small-scale quantum efficiency (QE) and pixel-size variations;
\item a collimated beam projector with a tunable laser \citep{2016SPIE.9910E..0VC}, to constrain wavelength dependence, nonuniform illumination, and scattered light in the dome flats;
\item an auxiliary telescope equipped with a low-resolution spectrograph, to constrain atmospheric transmission;
\item models fit to dithered observations of stars (``star flats''), to constrain degeneracies between QE and pixel size variation and provide an independent constraint on the illumination correction \citep[via methodology similar to that of][]{2017PASP..129k4502B}.
\end{itemize}
In ISR, our goal is to apply a flat that transforms the raw science image into one with \emph{surface brightness} pixels that is photometrically flat for objects with the spectral energy ditribution (SED) of the sky.
Such an image is inappropriate for precision photometry, but ideal for background subtraction.

The next few steps all fall into the category of single-epoch direct image characterization, in which we build models that describe the state of the observational system and how it transforms the true sky into the image that we see.
This includes background subtraction, PSF modeling, finding and interpolating over cosmic rays, measuring and applying aperture corrections, detecting, deblending, and measuring sources.
Each of these steps can only be done after at least one of the others, so in practice we'll have to repeat some of them as we iteratively improve the models they produce.

The detection, deblending, and measurement steps are essentially the same as those run in the SDSS \textit{Photo} pipeline \citep{2001ASPC..238..269L}:
\begin{itemize}
    \item Detection is responsible for identifying above-threshold regions and peaks within them; when multiple peaks appear in the same region, we call this a blend.
    \item Deblending creates a \emph{child} record for each peak in a blend, along with an image that contains the best estimate of the flux from just that child.
    \item Measurement applies a sequence of plug-in algorithms (e.g. centroiding, aperture photometry, PSF photometry, shapes) to both the \emph{parent} (the original image of each blend, hence interpreting it as a single object) and the child images produced by the deblender.
\end{itemize}
These are described much more fully in \citet{2018PASJ...70S...5B}.
As noted previously, we will probably need a new approach to deblending when processing deep (\textit{i.e.,} coadded) LSST images, but at single-epoch depths the SDSS algorithm still performs adequately.

Once we have produced a reliable source catalog, we match to a reference catalog (produced in the most recent data release\footnote{Plans for prompt processing prior to DR1 are still to be determined.}) to photometrically and astrometrically calibrate the image.
This first requires correcting the photometry for our background-optimized flat-fielding.
This could be done by dividing by our original flat-field image and multiplying by a new one that transforms to flux-valued pixels and flattens an SED more typical of astrophysical sources than the sky, prior to our final source measurement iteration.
In practice, we expect to be able to apply the same correction to sufficient accuracy in catalog-space after measurement.

\subsection{Image Differencing}

The heart of the Prompt pipelines is image subtraction and transient detection on the differences.
Each science observation is subtracted from a template coadd (also produced in the most recent data release) covering the same area of sky.
We first resample the template to the pixel coordinate system of the new science image, and then convolve it with a kernel to match its PSF to the PSF of the new observation.
Our baseline algorithm \citep{reiss_david_j_2016_192833} for computing this kernel is primarily based on that of \citet{1998ApJ...503..325A}, but incorporates some ideas from \citet{2016ApJ...830...27Z} as well.
The former maximizes the signal-to-noise of point-source detections in the limit that the template is noiseless, but does not rely on having accurate PSF models (which cannot be obtained in general in crowded fields).
The latter maximizes the signal-to-noise even when both images have noise but requires a high-quality Fourier-domain ratio of the template and science image PSFs as an input.
Our hybrid method should have the advantages of both approaches.

LSST lacks an atmospheric dispersion corrector, so differential chromatic refraction (DCR) makes the effective PSF (and hence the difference kernel) for each source a strong function of its SED.
This presents a significant challenge to all existing methods for image subtraction, which assume the difference kernel is a function only of position on the image.
Our baseline algorithm for dealing with DCR \citep{sullivan_ian_2018_1492936} involves constructing ``sub-band'' templates that can be combined with different weights for each source during image differencing.
The algorithm is still at a prototype stage, but early results are promising.

Once image subtraction is complete, we expect to run essentially the same detection algorithm we apply to single-epoch direct images to obtain \textit{DIASources} (Difference Imaging Analysis Sources).
Deblending on difference images is an easier problem than even deblending on shallow single-epoch direct images, because galaxies are static and should subtract cleanly, leaving only point sources, dipoles, and trailed sources that are significantly easier to model.
Many of our direct-image measurement algorithms can also be applied directly to difference images, though we will run additional measurement algorithms relevant only for moving objects.

Details of the measurements included in the \textit{DIASource} table (as well as all other LSST data products) can be found in the project's Data Product Definition Document \citep[LSE-163;][]{LSE-163}.

\subsection{Association and Alerts}

Before alerts are issued, new DIASources are spatially matched with the \textit{DIAObject} and \textit{SSObject} (Solar System Object) tables, providing extra context to include in the alert.
Entries in these tables are themselves constructed from unassociated DIASources (in the case of DIAObject) and sets of linked DIASources that are consistent with solar system orbits (SSObjects).
New DIAObjects are created and existing DIAObjects are updated immediately after DIASource measurement, while SSObject linkage and orbits are re-analyzed during the 24-hour period following observation.

We also perform \emph{precovery} on all new DIAObjects by performing forced photometry at their positions in all difference images observed in the previous 30 days, in order to provide light-curves for new transients before they rise above our detection threshold.
Precovery is also run during the 24 hours following the observation of the object.

Alert packets will be sent out 60 seconds after each observation for all DIASource measurements obtained in that observation.
The packets will include the DIASources, their associated DIAObjects or SSObjects, any associated DIASources from the last 12 months, and identifiers for several nearby Objects from the most recent Data Release Processing.
The Prompt Products Database (PPDB) holding the DIASource, DIAObject, and SSObject tables will also provide an interface to this information; it will be queryable about 24 hours after the observation.

\section{Data Release Pipelines and Data Products}

\subsection{Image Characterization and Calibration}

The Data Release pipelines begin with essentially the same single-epoch processing that is run in the Prompt pipelines.
Each LSST Data Release is intended to be entirely independent, however, so the LSST-produced reference catalogs and templates used in Prompt processing are not an option.
Instead, we plan to match and calibrate to the Gaia catalog \citep{2016A&A...595A...1G}, but in the Data Release pipelines this just provides a preliminary calibration.
For the final calibration, we will move away from processing each observation independently and instead fit models all single-epoch catalogs in each area of sky together.

For the astrometric calibration, we will utilize a joint fit similar to that of \citet{2017PASP..129g4503B} to constrain the instrumental degrees of freedom as well as a simple empirical (e.g. polynomial) model for astrometric offsets introduced by the atmosphere.
This fit will be constrained to exactly reproduce the positions of stars measured by Gaia, which obtains much better astrometric accuracy on bright stars than LSST can hope to achieve; our goal here is essentially to extend the astrometric calibration to smaller spatial scales generally unconstrained by the shallower and hence sparser Gaia catalog.

The details of LSST's photometric calibration are still somewhat uncertain, depending in part on the ultimate photometric precision of the Gaia catalog.
Joint modeling of multiple observations of the same objects will play a major role, just as in astrometric calibration, and we are currently  integrating the forward-modeling approach used to calibrate the Dark Energy Survey \citep{2018AJ....155...41B} into our pipelines to extend this to physically-motivated modeling of the full photometric system.
In the future, this will also incorporate the collimated beam projector data and auxiliary telescope spectra mentioned previously.

After joint calibration, we will return to single-epoch processing, however, to fit an improved PSF model by utilizing the more precise astrometric calibration and more secure star/galaxy classifications provided by the joint analysis of multiple epochs.

\subsection{Coaddition and Image Differencing}

At this point, our single-epoch images are almost fully characterized, but we expect to be able to improve on both our sky background model and our understanding of which pixels are affected by artifacts (e.g. cosmic rays, satellite trails, and optical glints and ghosts) via an image differencing analysis similar (but not identical) to that run in the Prompt pipelines.

This is particularly clear for artifact masking, because most artifacts will only appear in a single epoch, and even optical ghosts due to bright stars will appear in different positions as long as observations are dithered.
The argument for background modeling is more subtle; it hinges on the fact that one of the biggest challenges in background estimation is separating variation in the sky from ``astrophysical'' backgrounds that we would like to preserve, such as diffuse light from galaxy clusters.
Very few of these astrophysical backgrounds are time-variable, however, so they should subtract cleanly in image subtraction.
Given $N$ single-epoch images that cover some area of sky, then, we can construct difference image pairs that fully constrain $N-1$ sky backgrounds.
That leaves one ``reference'' image with its sky background intact and confused with the (common) astrophysical background.
By combining the sky-subtracted images with the reference image to build a deeper coadd, we can enhance the signal-to-noise of the astrophysical background relative to the single remaining sky background, making them easier to separate in the final subtraction.

As we build coadds (including templates) and use them in image differencing analysis, we consequently also improve our background models and artifact masks.
Just like the steps in initial characterization, the detailed ordering of the subsequent few processing steps is still to be determined due to circular dependencies that we must iteratively "unroll".
These steps include interpolating masked pixels, resampling all single-epoch images in each patch of sky to a common pixel grid, and comparing the resampled images to improve artifact masks and constrain sky-background pairs.

Once this iterative processing is complete, we can build final coadds and subtract the final background.
This includes constructing template coadds to be used in image differencing in both the current Data Release and future Prompt processing; at this point we run essentially the same code in both productions, at least through DIASource production.

We also now return to single-epoch direct-image processing to produce the \textit{Source} table, by performing one more round of detection, deblending, and measurement on the now fully-characterized single-epoch images.

\subsection{Object Detection and Measurement}

Entries in the \textit{Object} table represent our best measurements of each astrophysical object in a given Data Release, and as a whole the table serves as a hub for other table data products.
Objects can be produced in two ways: from coadd detections and DIAObjects (\textit{i.e.,} DIASource associations).
Sources derived from \emph{direct} single-epoch detections are \emph{not} directly associated into Objects; any astrophysical object should already appear in either coadd detections or DIASources (and be better characterized in at least one of those).

Detection on coadds will use the same algorithms employed on single-epoch direct (producing Sources) and difference images (producing DIASources), with the added complication that here we need consistent detections across all bands.
We can achieve this either by combining coadds from different bands prior to detection \citep[as in][]{1999AJ....117...68S}, or by detecting on the coadd for each band separately and merging them in catalog-space.
We expect to use both approaches at some level (\textit{i.e.,} detecting on different combinations of bands and merging the results).
We may also create and detect on coadds created from inputs restricted to limited observation date ranges, in order to improve our ability to detect slowly-moving faint objects.

In Prompt processing, new DIASources are associated with and used to update living DIAObject and SSObject tables.
Because we start from scratch in each Data Release, there we instead associate all DIASources in each patch of sky to immediately create complete DIAObjects and SSObjects.
DIAObjects are intended to represent astrophysical objects that either do not move or move very little, and hence each DIAObject is either associated with an existing Object derived from a coadd detection or used to create a new Object record.
Because SSObjects appear in a very different place every time they are observed, they are not included in the Object table.
Whenever possible, we hope to have different Objects to reflect astrophysically distinct but spatially coincident entities, including distinguishing galaxy Objects from any time-variable AGN or supernovae they host, by utilizing time-domain information as well as positional information in the association.

Constructing secure Object definitions is probably impossible without adding morphological information into the mix, and this blurs the line we have drawn between detection and deblending in both the previous steps and in the current implementation of the pipeline.
The Object deblending algorithm is very much still a work in progress, but the Scarlet algorithm \citep{2018A&C....24..129M} is a likely starting point, and we can say that the full algorithm will utilize coadds from all bands simultaneously as well as DIAObject information, and it may modify the preliminary Object definitions passed to it from detection and association.

As in all deblending contexts, the Object deblender is also responsible for creating deblended child images.
These are then used for measurement on coadds in approximately the same manner as in single-epoch processing, with the main differences being the set of algorithms that are run and the fact that at least some of these will utilize images from multiple bands simultaneously.

Two measurement algorithms will go considerably further, and fit models simultaneously to all of the original single-epoch images that overlap the Object.
The first of these is a stellar astrometry model that fits a point source with proper motion and parallax parameters, which cannot be fit to coadds at all.
Compared to traditional astrometry methods that fit motion parameters to independently-measured source positions, this approach \emph{should} enable us to extend measurements to fainter magnitudes and better handle blending with faint neighbors \citep{2009AJ....137.4400L}.

The second model is a two-component PSF-convolved galaxy model, intended to approximate a bulge-disk decomposition (a full bulge-disk decomposition would require more signal-to-noise and resolution than the vast majority of LSST galaxies will have).
At some level, this model could be fit only to the coadds, which should generally provide a good representation of the static sky.
Utilizing the single-epoch pixel data makes it easier to guarantee that we have maximized the signal-to-noise of the measurement and avoided systematic errors (but only if all relevant single-epoch images are included in each likelihood evaluation).
We are investigating approaches to building coadds of sufficient quality that going back to single-epoch images at this stage would be unnecessary, which would be much more computationally efficient.
Systematic errors are the biggest concern, particularly those related to the print-through of CCD edges onto the coadd and noise correlations due to resampling.
Methods exist to avoid signal-to-noise loss \citep[e.g.][]{2011PASP..123.1117H,2017ApJ...836..188Z} in coaddition, but even without these the typical losses (which depend on the seeing distribution) seem to be minimal \citep{2018PASJ...70S...5B}.

Finally, we will return to both single-epoch direct images and difference images to perform forced photometry at the position of each Object.
This should provide our best estimates of the light-curves of Objects, including Objects that were never detected in single-epoch images.
We are currently planning to run this on both direct and difference images because direct images may provide slightly better signal-to-noise while difference images should provide much better control over systematics due to deblending.
Deblending in single-epoch direct forced photometry -- in which most blend children are not even detectable -- is at some level impossible, and we generally expect difference-image forced photometry to provide better results overall, simply because blends should be much rarer and much less severe in difference images.

\section{Conclusion}

The LSST pipelines today are already representative of the state of the art in large-scale optical image processing.
The future pipelines described here go considerably beyond this both in scale and in algorithmic sophistication, and there is hence substantial work to be done and a very good chance that some our of approaches to these problems will change before LSST first light.
These challenges are present in both Prompt and Data Release processing, as well as in integrating both of these with each other and with the rest of the operational system.

Pushing the boundaries in software as well as hardware is a matter-of-course for new scientific projects, of course, and good progress is certainly being made.
When operational, the LSST pipelines will produce catalogs that enable similarly boundary-pushing science, in most cases without requiring further access to the pixel data.

\acknowledgements
We thank Keith Bechtol for helpful comments on earlier drafts of this paper.
This material is based upon work supported in part by the National Science Foundation through Cooperative Agreement 1258333 managed by the Association of Universities for Research in Astronomy (AURA), and the Department of Energy under Contract No. DE-AC02-76SF00515 with the SLAC National Accelerator Laboratory.
Additional LSST funding comes from private  donations, grants to universities, and in-kind support from LSSTC Institutional Members.

\bibliography{I12-1}

\end{document}